\documentclass[12pt]{iopart}
\usepackage{graphicx}

\begin{document}

\title{Identified Hadrons and Jet Chemistry for p+p and Au+Au Collisions at RHIC}

\author{Yichun Xu for the STAR collaboration }

\address{Department of modern physics, University of science and technology of China, Hefei, Anhui 230026, China}\ead{xuyichun@mail.ustc.edu.cn}

\begin{abstract}


The study of hadron spectra at high $p_{T}$ in p+p collisions
provides a good test of perturbative quantum chromo-dynamic
calculations (pQCD) and a baseline for measurements of nuclear
modification factors in Au+Au collisions. Using events triggered
with electro-magnetic calorimetery, identified charged
hadron transverse momentum ($p_T$) spectra are measured up to 15
GeV/$c$ at mid-rapidity ($\mid y\mid$ $<$ 0.5) and neutral kaon
$p_T$ spectra up to 12 GeV/$c$ in p+p collisions at
$\sqrt{s_{NN}}$ = 200 GeV in the STAR Experiment~\cite{STAR}.
The particle ratios of $p/\pi^{+}$,
$\overline{p}/\pi^{-}$ and $K^{\pm,0}$/$\pi^{\pm}$ in p+p
collisions are shown and compared with next-to-leading order pQCD
calculations. In central Au+Au collisions, we report nuclear
modification factors ($R_{AA}$) for pion, kaon, proton and $\rho$
and discuss several model calculations: color-charge dependence of
jet quenching and jet conversion. Finally, centrality dependence
of $R_{AA}$ at high $p_T$ ($>$ 5.5 GeV/c) for kaons are compared
with that of pions in Au+Au collisions at 200 GeV.
\end{abstract}


\section{Introduction:}
Measurements of identified charged hadron ($\pi^{\pm}$, $K^{\pm}$,
$p(\overline{p}$)) transverse spectra at $p_T\!>2$ GeV/$c$ in
elementary p+p collisions provide a good test of perturbative
Quantum Chromodynamics (pQCD) \cite{pQCD}. In pQCD calculations, the
inclusive production of single hadrons is described by the
convolution of parton distribution functions (PDF), parton-parton
interaction cross-sections, and fragmentation functions (FF). The
FFs were parameterized according to experimental measurements in
$e^{+}e^{-}$ collisions but are not well constrained. A more precise
constraint on the quark and gluon FFs by comparing theory with
experimental data is crucial to understand hadron production
mechanisms. In addition, these measurements in p+p collisions at
high $p_T$ provide an essential baseline for studying parton energy
loss in heavy ion collisions. We use events triggered by an
electro-magnetic calorimeter and extend the identified charged
particle measurements up to 15 GeV/$c$, and neutral kaons up to 12
GeV/$c$ in p+p collisions at $\sqrt{s_{NN}}$ = 200 GeV. This
significantly extends particle $p_T$ reach beyond previously
published measurements of up to 10, 5, and 7 GeV/$c$ respectively
for pions, $K_S^0$, and protons in p+p collisions using events
triggered with minimum bias~\cite{starppPID,starppKs}. The
relativistic ionization energy loss ($dE/dx$) in the Time Projection
Chamber (TPC) was used for charged hadron identification. A new
method to re-calibrate the TPC $dE/dx$ was developed and
applied~\cite{re-cali}, which significantly reduced the systematic
uncertainties for protons compared to previous
measurements~\cite{starppPID,starAuAuPID}. In these proceedings,
identified particle $p_T$ spectra, and the particle ratios
($p/\pi^+$, $\overline{p}/p$ and $K^{\pm,0}/\pi^{\pm}$) are
presented and compared with next-to-leading order (NLO) pQCD
calculations, and with PYTHIA results. Nuclear modification factors
$R_{AA}$, the spectra in Au+Au collisions divided by spectra in p+p
collisions scaled by the number of binary collisions, are presented
and compared with theoretical calculations~\cite{jetConversion}.

\section{Experiment and Analysis}
The p+p collision events with enhanced high-$p_T$ charged
particles we use in this analysis are obtained from an online trigger
by the Barrel Electro-Magnetic Calorimeter (BEMC) with
0$<$ $\eta$ $<$ 1 and full (2$\pi$) azimuthal coverage in
the year 2005~\cite{BEMC}. Charged particle tracking is performed
with the Time Projection Chamber (TPC), spanning $|\eta|$ $<$ 1.8
and 2$\pi$ in azimuth, which enables particle identification through
measurements of momentum and $dE/dx$~\cite{tpc}. In p+p
collisions, $\sim$5.6 million events triggered by
transverse energy $E_{T}$ $>$ 6.4 GeV in single BEMC towers
(JP2 events), $\sim$5.1 million events with $E_{T}$ $>$ 2.5
GeV (HT1 events), and $\sim$3.4 million events with $E_{T}$ $>$ 3.6 GeV
(HT2 events) are used for the analysis.
Neutral kaons are reconstructed in the HT1 and HT2 triggered events
through the $K^{0}_{S} \rightarrow \pi^{+} + \pi^{-}$ decay mode, while
JP2 triggered events are used to measure charged hadrons. These events
also serve to check for a trigger bias by comparing $K^{0}_{S}$ on the
near side of the trigger (in azimuth) with $K^{\pm}$ on the away side
using other triggered events. The charged kaon analysis involves $\sim$21.2
million central Au+Au collisions.

\begin{figure}\centering
\includegraphics[width=0.45\textwidth]{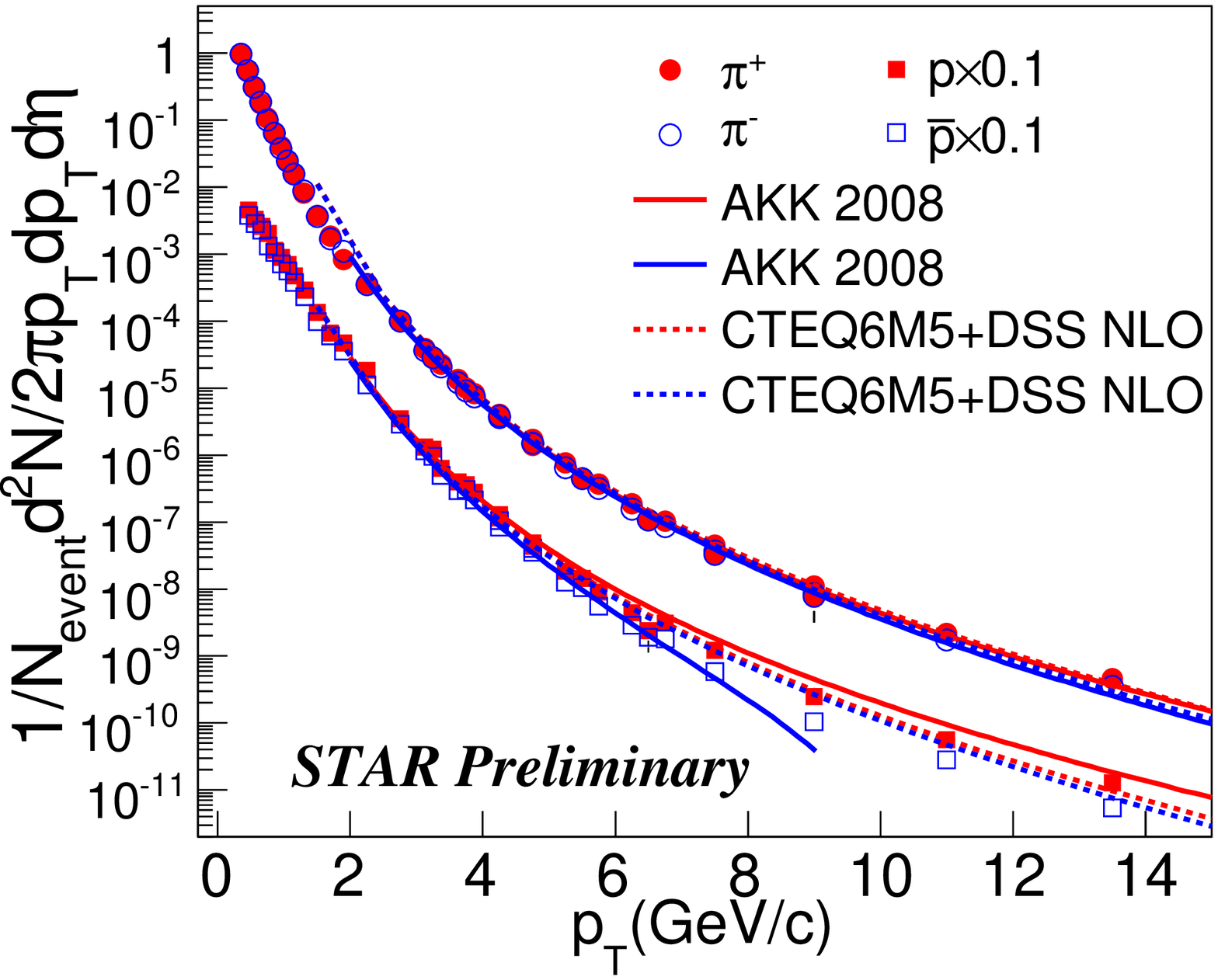}
\includegraphics[width=0.45\textwidth]{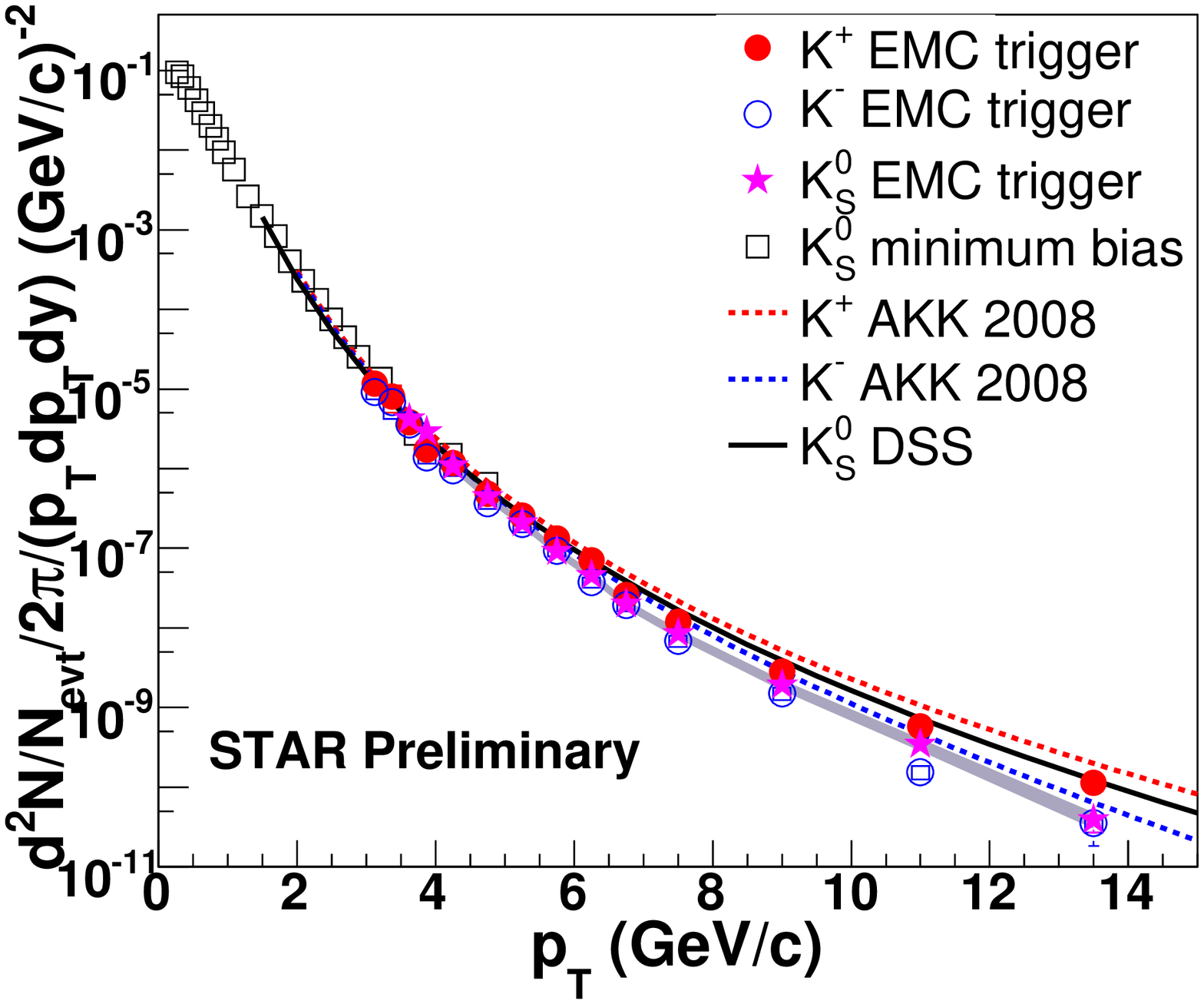}
\caption[]{Charged hadrons and neutral kaon $p_T$ spectra in p+p
collisions, compared with pQCD NLO calculations from DSS, AKK 2008.
The shaded band on the right panel represents systematic
uncertainties for $K^0_S$.} \label{ppPID}
\end{figure}

The segment length $x$ dependence of $dE/dx$, gas multiplication
gains, noise of TPC electronics, and pileup in high luminosity
environment may cause the measured $dE/dx$ of charge particles in
the TPC to deviate from the expected values, as calculated from the
Bichsel function~\cite{Bichsel}. In the relativistic rise region,
the separation of $dE/dx$ peaks among $\pi^{\pm}$, $K^{\pm}$ and
p($\overline{p}$) are between approximately 1 to 3 $\sigma$. Pions
are the dominant particle species for inclusive and jet hadrons, and
their $dE/dx$ distributions overlap the smaller kaon and (even
smaller) proton distributions within any given momentum slice,
preventing clear peak separation. This results in large systematic
errors in attributing yields to species due to the uncertainty of
the reconstructed $dE/dx$ peak positions. The re-calibration method
is important to improve these systematics~\cite{re-cali,HotQuark}.
With the Bichsel function, normalized $dE/dx$ distributions for
charged particles can be fit by an 8-Gaussian function to obtain
identified particle yields at given momenta~\cite{ming,re-cali}. To
correct for the trigger enhancement, PYTHIA is used to generate
events which are then passed through GEANT, and the fraction of
those events passing the trigger threshold is determined. The
combined acceptance and efficiency correction is found to be 88$\%$
from GEANT simulations. Final spectra for charged hadrons at
$|\eta|$ $<$ 0.5 are shown oi Fig.~\ref{ppPID} and are consistent
with previously published spectra in minimum bias
events~\cite{starppPID}.

For the trigger bias cross check, near-side $K^{0}_{S} \rightarrow \pi^{+} + \pi^{-}$
decays are reconstructed in HT1 and HT2 triggered events
through identified displaced vertices (V0s)~\cite{starppKs}.
We require one of the daughter pions to have fired the
BEMC tower energy trigger and compare the $p_T$ spectra of
these pions to those from minimum bias events to determine
trigger efficiency versus $p_T$. Simulations provide a combined
efficiency of 55.6\% for reconstructing the second daughter pion
and identifying the V0.
Final $K^{0}_{S}$ $p_T$ spectra are shown as stars on the
right panel of Fig.~\ref{ppPID}. Within uncertainties,
they are consistent with charged kaon spectra,
indicating sufficient removal of trigger biases by our corrections.
The NLO pQCD calculations (AKK 2008~\cite{AKK2008} and
DSS~\cite{DSS}) shown in Fig.~\ref{ppPID} exhibit good agreement
with charged pion spectra, but cannot describe our (anti-)proton and
kaon spectra~\cite{HotQuark,QM09}. Our data can be used to further
constrain parameters of FFs in these calculations.

\begin{figure}\centering
\includegraphics[width=0.45\textwidth]{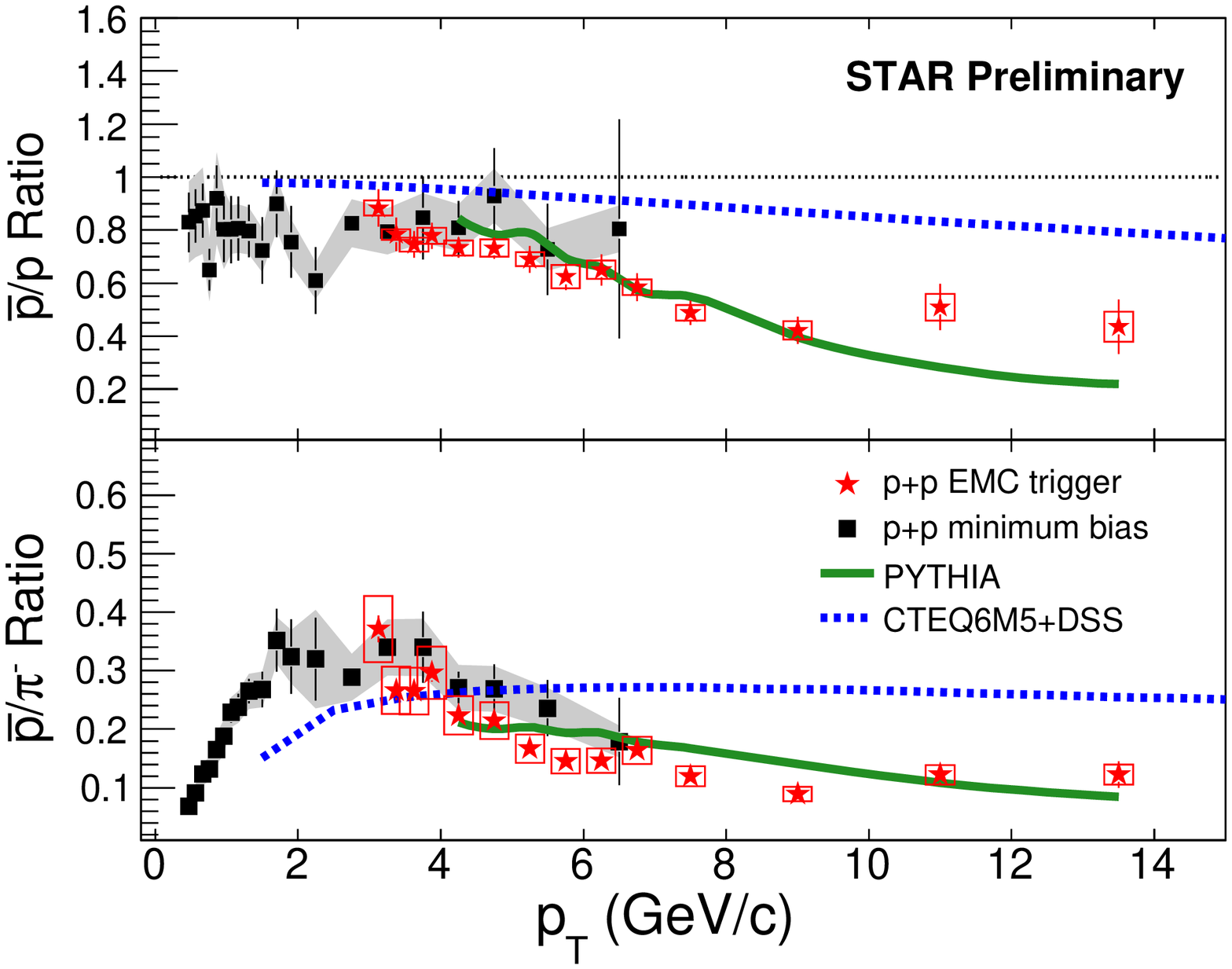}
\includegraphics[width=0.5\textwidth]{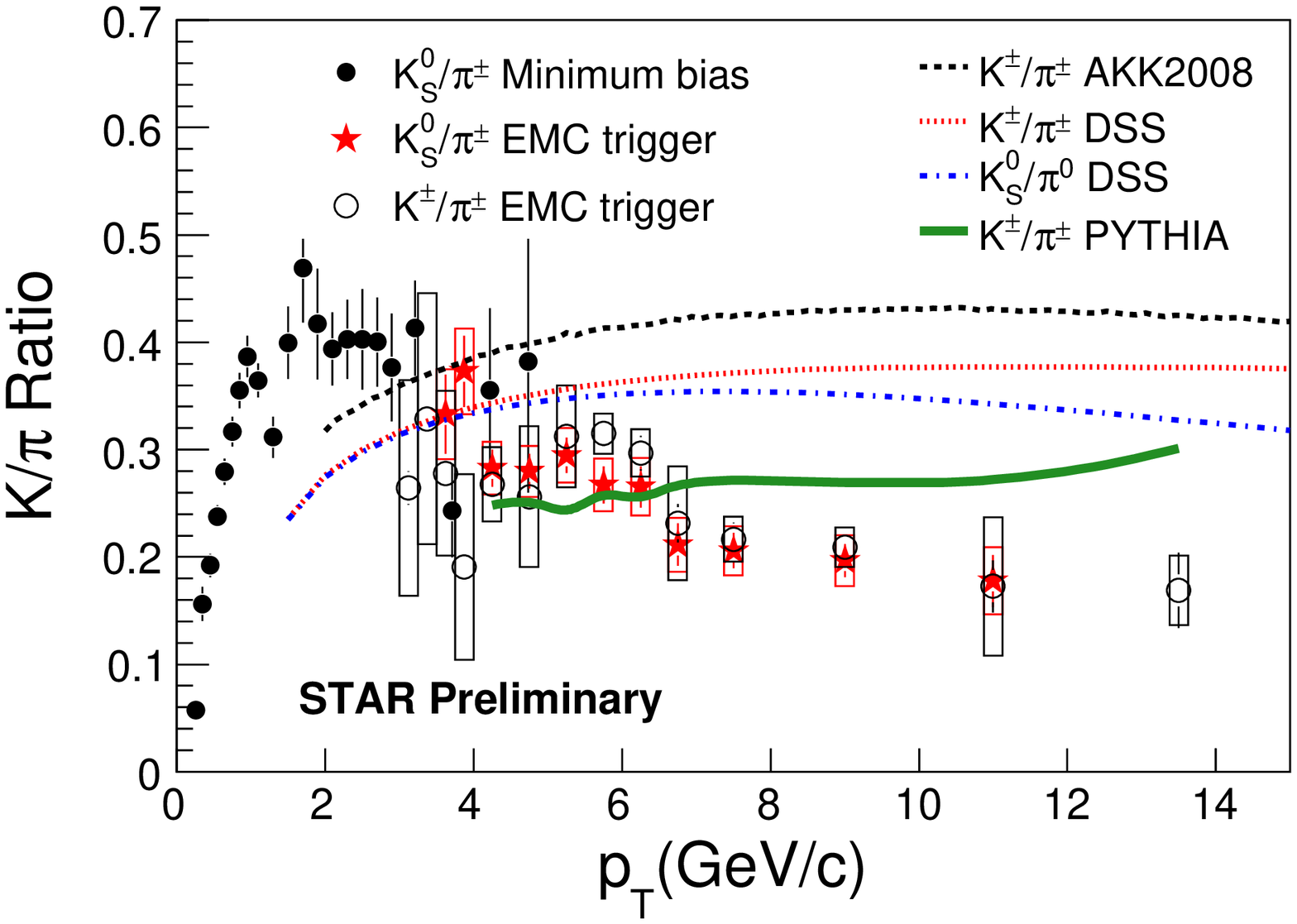}
\caption[]{Experimental $\overline{p}$/p, $\overline{p}$/$\pi^{-}$
(left), and K/$\pi$ (right) ratios in p + p collisions, compared
with pQCD NLO calculations from DSS and PYTHIA. Vertical lines show
statistical errors, while shading and boxes represent systematic
uncertainties.
} \label{ppRatio}
\end{figure}

In addition, Fig.~\ref{ppRatio} shows several particle ratios
($\overline{p}$/p, $\overline{p}/\pi^{-}$, and
$K^{\pm,0}$/$\pi^{\pm}$)~\cite{HotQuark,QM09} as a function of $p_T$
at mid-rapidity in the BEMC triggered events from p+p collisions
along with published results from minimum bias p+p
collisions~\cite{starppPID,starppKs} and predictions from pQCD
calculations (DSS NLO calculations and PYTHIA simulations).
The experimental data are consistent in regions of overlapping $p_T$
and show minor divergences from PYTHIA, but are well below
the predictions of DSS at high
$p_T$. This again show that our data can provide a good constraint
to the NLO pQCD calculations.

\begin{figure}
\centering
\includegraphics[scale=0.5]{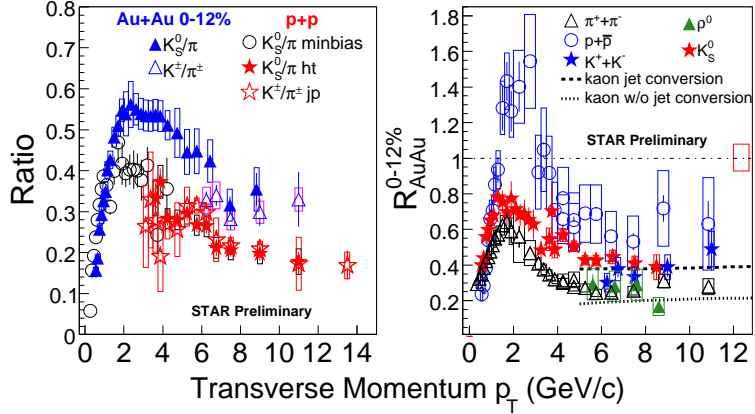}
\caption{K/$\pi$ ratios and nuclear modification factors of $\pi$,
$K$, $p$ and $\rho$ in Au+Au collisions as a function of $p_T$.
The bars and boxes represent statistical and systematic
uncertainties respectively.} \label{Raa}
\end{figure}

Charged particles are identified in Au+Au collisions via the same
re-calibrated dE/dx method~\cite{re-cali}. The K/$\pi$
ratios in Au+Au collisions are shown in
Fig.~\ref{Raa} (left) compared with the ratios in p+p collisions.
The enhancement of K/$\pi$ in Au+Au collisions demonstrates less
suppression for kaons than pions at high $p_T$. To further
understand this phenomenon, $R_{AA}$ in central Au+Au collisions
are shown in Fig.~\ref{Raa} (right) for kaon, pion,
proton, and $\rho$~\cite{rhoQM08}. The study of detailed systematic
uncertainties for kaons and protons is underway. We observe that
$R_{AuAu}(K^{\pm},K^{0}_{S})$ is larger than $R_{AuAu}(\pi^{\pm})$,
which is in contradiction to the prediction from a model based on
parton energy loss through gluon radiation~\cite{energyLoss}.
That $R_{AuAu}(\pi^{\pm})$ is similar to
$R_{AuAu}(\rho)$ indicates that the difference is not a mass
effect. The value of $R_{AuAu}(K^{\pm},K^{0}_{S})$ is
consistent with the prediction from jet conversion in the hot dense
medium (dashed line)~\cite{jetConversion}. The same
factor, scaling the lowest-order QCD jet conversion rate, applied to
calculate proton $R_{AA}$ \cite{starAuAuPID,Liu} is used in this
prediction.

\begin{figure}
\centering
\includegraphics[scale=0.3]{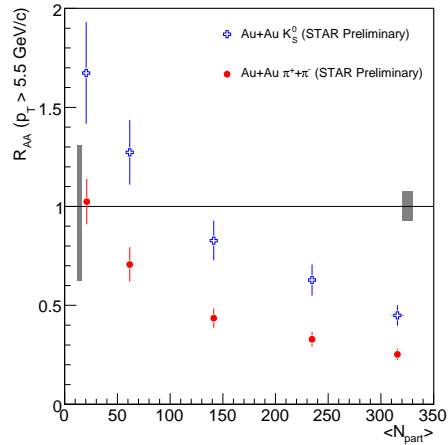}
\caption{Integrated $R_{AA}$($K^0_S$) and $R_{AA}$($\pi$) for $p_T$
$>$ 5.5 GeV/$c$ in Au+Au $\sqrt{s_{NN}}$ = 200 GeV collisions. The
uncertainties on the data points are statistical and systematic
added in quadrature. The left and right
grey bands represent the typical uncertainties on $<$$N_{bin}$$>$ for
peripheral and central Au+Au $\sqrt{s_{NN}}$ = 200 GeV collisions
respectively.} \label{Raavscent}
\end{figure}

In order to understand more about the jet chemistry change in the
medium, we also show the centrality dependence of $R_{AA}$ in
Fig.~\ref{Raavscent}. Due to limited statistics, we plot integrated
$R_{AA}$ for $p_T$ $>$ 5.5 GeV/$c$ (with mean $<$$p_T$$>$ $\sim$ 6.2
GeV/$c$ in this range), where particle production may be dominated by
hard processes.

Both kaon and pion production are suppressed in central collisions,
and $R_{AA}(K)$ is about a factor of 2 larger than $R_{AA}(\pi)$,
even in peripheral collisions. This raises the question whether
parton flavor conversion is prevalent, even in the smaller systems,
or if there is some other soft production mechanism than jet
fragmentation contributing in this $p_T$ range for all centralities
of A+A collisions.

\section{Summary and discussion}
Using the BEMC triggered events and a new $dE/dx$ re-calibration
method, $\pi^{\pm}$, $K^{\pm}$ and $p(\overline{p}$) transverse
momentum ($p_T$) spectra in p+p collisions at $\sqrt{s_{NN}}$ = 200 GeV
are extended up to 15 GeV/$c$ at
mid-rapidity ($|y|$ $<$ 0.5), and neutral kaon $p_T$ spectra up to 12
GeV/$c$. The
ratios $p/\pi^{+}$, $\overline{p}/\pi^{-}$, and
$K^{\pm,0}$/$\pi^{\pm}$ in p+p collisions are shown and compared
with NLO pQCD calculations. These
calculations cannot reproduce kaon and proton spectra at high
$p_T$. In central Au+Au collisions,
we observe $R_{AA}(p+\overline{p})\geq
R_{AA}(K^{0}_{S},K^{\pm})>R_{AA}(\pi^{+}+\pi^{-})\simeq
R_{AA}(\rho^{0})$ at high $p_T$ (above 5.5 GeV/$c$).
Jet conversion together with other mechanisms,
such as parton splitting~\cite{split}, might be able to explain these
observations. Higher statistics from Run 10 may give us additional
physics information from more precise measurements of the nuclear
modification factors.

\section{Acknowledgments}
This work was supported in part by the National Natural Science
Foundation of China  under Grant No. 10610286(10610285), 10475071,
10805046 and the Knowledge Innovation Project of Chinese Academy of
Sciences under Grant No. KJCX2-YW-A14.


\section*{References}
\begin{thebibliography}{10}

\bibitem{STAR}
K.H. Ackermann {\it et al.}. Nucl. Instr. and Meth. A 499 (2003) 624.

\bibitem{pQCD}
J.C. Collins, D.E. Soper, Annu. Rev. Nucl. Part. Sci. 37 (1987) 383;

\bibitem{starppPID}
J. Adams {\it et al.}, Phys. Lett. B 637 (2006) 161-169; J. Adams
{\it et al.}, Phys. Lett. B 616, 8 (2005).

\bibitem{starppKs}
  B.~I.~Abelev {\it et al.}  [STAR Collaboration],
  Phys.\ Rev.\  C {\bf 75}, 064901 (2007)
  [arXiv:nucl-ex/0607033].

\bibitem{re-cali}
Y. Xu {\it et al.}, Nucl. Instr. and Meth. A 614 (2010) 28-33,
arXiv:0807.4303.

\bibitem{starAuAuPID}
B.I. Abelev {\it et al.}, Phys. Rev. Lett. 97 (2006) 152301; B.I.
Abelev {\it et al.}, Phys. Lett. B 655, 104 (2007).

\bibitem{jetConversion}
W. Liu, R.J. Fries, Phys. Rev. C77 (2008) 054902.

\bibitem{BEMC}
M. Beddo {\it et al.} Nucl. Instr. Meth. A 499 (2003) 725-739.

\bibitem{tpc} M. Anderson {\it et al.}, Nucl. Instr. Meth. A 499 (2003) 659.

\bibitem{Bichsel}
H. Bichsel, Nucl. Instr. Meth. A 562 (2006) 154.

\bibitem{HotQuark}
  Y.~Xu [STAR Collaboration],
  Eur.\ Phys.\ J.\  C {\bf 62}, 187 (2009)
  [arXiv:0901.0692 [nucl-ex]].

\bibitem{ming}
M. Shao {\it et al.}, Nucl. Instr. Meth. A 558, 419 (2006).

\bibitem{AKK2008}
  S.~Albino, B.~A.~Kniehl and G.~Kramer,
  Nucl.\ Phys.\  B {\bf 803}, 42 (2008)
  [arXiv:0803.2768 [hep-ph]].


\bibitem{DSS}
  D.~de Florian, R.~Sassot and M.~Stratmann,
  Phys.\ Rev.\  D {\bf 75}, 114010 (2007)
  [arXiv:hep-ph/0703242].

\bibitem{QM09}
  Y.~Xu  [STAR Collaboration],
  Nucl.\ Phys.\  A {\bf 830}, 701C (2009)
  [arXiv:0907.4644 [hep-ph]].

\bibitem{rhoQM08}
Patricia Fachini, Acta Phys.Polon. B35 (2004) 183-186

\bibitem{energyLoss}
X. N. Wang, Phys. Rev. C 58, 2321 (1998).

\bibitem{Liu}
W. Liu, C.M. Ko and B.W. Zhang, nucl-th/0607047.

\bibitem{split}
S. Sapeta and U.A. Wiedemann, arXiv:0707.349.



\end{thebibliography}
\end{document}